\documentclass[journal]{IEEEtran}

\usepackage[T1]{fontenc}
\usepackage[utf8]{inputenc}

\usepackage{cite}
\usepackage{amsmath,amssymb,amsfonts}
\usepackage{algorithm} 
\usepackage{algorithmic}
\usepackage{graphicx}
\usepackage{textcomp}
\usepackage{xcolor}
\def\BibTeX{{\rm B\kern-.05em{\sc i\kern-.025em b}\kern-.08em
    T\kern-.1667em\lower.7ex\hbox{E}\kern-.125emX}}
\makeatletter
\newcommand{\removelatexerror}{\let\@latex@error\@gobble}
\makeatother

\newcommand{\queschain}{Ques-Chain}

\begin{document}
\title{\queschain: an Ethereum Based E-Voting System 
}

\author{\IEEEauthorblockN{Qixuan Zhang, 
Bowen Xu, 
Haotian Jing
and Zeyu Zheng}\\
\IEEEauthorblockA{
School of Information Science and Technology\\
ShanghaiTech University, China\\
\texttt{\{zhangqx1, xubw, jinght, zhengzy\}@shanghaitech.edu.cn}}
}

\maketitle

\begin{abstract}
    Ethereum is an open-source, public, blockchain-based distributed
    computing platform and operating system featuring smart contract
    functionality. In this paper, we proposed an Ethereum based 
    eletronic voting (e-voting) 
    protocol, \queschain{}, which can ensure the authentication can be done
    without hurting confidentiality and the anonymity can be protected
    without problems of scams at the same time. Furthermore, the authors
    considered the wider usages \queschain{} can be applied on, pointing out
    that it is able to process all kinds of messages and can be used in all
    fields with similar needs.
\end{abstract}

\begin{IEEEkeywords}
    Electronic voting, Ethereum, Smart contracts, Blind signature
\end{IEEEkeywords}

\section{Introduction}

E-voting and web-based survey are popular ways to collect opinions from
citizens, clients, employees and sometimes organizations or companies.
The conductors of elections, evaluations or questionnaires expect voters
to give real and practical comments on their particular topics. However,
e-voting systems face the threat of malicious manipulation by hackers.
Questionnaire and poll service providers also struggle to prevent and
eliminate scams, given the high costs to perform data cleaning. On the
other hand, voters may doubt the integrity of the voting procedure and
worry about anonymity failure. Thus, an e-voting system ought to be
capable of conducting authentication, providing transparency, protecting
anonymity, securing ballots, and yielding accurate statistics.
Blockchain, initially introduced as a distributed ledger and nowadays a
computation vender, has proved its inherence in providing immutability,
verifiability and decentralized consensus throughout its very first
application, bitcoin the cryptocurrency. Ethereum (ETH), a blockchain, and
Smart Contract, a way to assign computing missions to the Ethereum
network, have become the de facto standard of blockchain-based trusted
computing. Provided the outstanding features, it is promising to base
future e-voting systems on blockchain technology.

The paper is structured as follows: section \ref{releated_work} introduces existing works
on e-voting and blockchain, section \ref{contributions} sums up our main contributions,
section \ref{techniques} presents two techniques which our work based on. \queschain{}'s
mathematical model and scheme detail are provided respectively in
section \ref{notations} and section \ref{details}. \queschain{}'s security properties are analyzed
in section \ref{security}. Section \ref{usages} introduces some application scenarios of
\queschain{}.

\section{Related Work}\label{releated_work}

There has been extensive research on adopting blockchain into e-voting
and surveys. Reference \cite{pawlak2018towards} introduced a 
multi-agent system, in which
several kinds of intelligent agents act as blockchain computing nodes,
to ensure voters' right to audit and verify the voting process. Liu and
Wang designed a feasible and flexible e-voting scheme
with no dependency on time-triggered protocol (TTP), 
trusted third party, on the blockchain and fulfilled criteria on
general voting systems \cite{liu2017voting}. 
Liu and Wang also provided certain remedies for
data neutrality deficiency and privacy risk in data transmission. Panja
and Roy applied blockchain technology to the existing DRE-ip e-voting
system, which protects verified ballots from being modified before the
tallying phase and provide a substitute 
to secure bulletin board \cite{panja2018secure}.
Permissioned sidechains \cite{francesco2018crypto} can be adopted in e-voting
systems for voter verification and voting operation recording, each
computed and stored on its respective one-way pegged sidechain but
linked by the upper layer network. Hjalmarsson et al. conducted votes on
private chains with the utilized smart contract to improve the
processing speed and throughput, as well as to ease the load and save
computational spends \cite{hjalmarsson2018blockchain}. 
Blockchain-based computing networks, by exploiting
the consensus of nodes, are able to reject fraudulent ballots and
guarantee the voting result unforgeable 
and transparent \cite{hardwick2018voting}.

\section{Main Contributions}\label{contributions}
The main contributions of this paper are summed up as follows:
\begin{enumerate}
\item
  We have proposed a message authentication and transmission mechanism
  that allows permission checking while preserving anonymity. The
  mechanism can be utilized in various scenarios including vote,
  questionnaire, outcome assessment, opinion collection, complaint
  reporting and so on.
\item
  We have decoupled the blind signing and checking procedure into three
  steps, respectively processed by the organizer, the voters, and the
  \queschain{} smart contract. With such a design, authentications can be
  carried out without sacrificing voters' anonymity and messages'
  confidentiality. In applications like opinion collection, it can also
  prevent spamming.
\item
  We have implemented the complete mechanism, featuring trusted
  computing technology based on Ethereum. Our well-designed architecture
  guarantees the integrity of all parties.
\end{enumerate}

\section{Main Techniques}\label{techniques}
In this session, we will introduce main techniques we used in our
protocol.

\subsection{Blind Signature}
Blind signature was firstly introduced by Chaum\cite{chaum1983blind}. Same as digital
signature, blind signature is used for validating the authenticity of a
message. The difference between both methods of signature is that the
messages are blinded and encrypted, which means that the data being
signed varies from the original message. And the method assure that the
one who receive the message can get the original message by unblinding
and decrypting it. The process is same as applying a legacy digital
signature to the original message. Therefore, the authenticity of the
message can be verified by validating it with the public key of the
signature. Blind signature can be used in e-voting system to perform
better privacy protection for voters \cite{ibrahim2003secure}.

\subsection{Ethereum}
Ethereum acts as a general decentralized computing platform based on
blockchain for economic benefit and new kinds of calculating
applications. It offers a decentralized Turing complete virtual machine
called Ethereum virtual machine (EVM), where scripts for the platform can be
run. The scripts are called smart contracts, which is mostly written in
Solidity, a script language designed for EVM. 
After deployment, the
scripts will automatically execute in decentralized network
transparently. Transaction and deployment in Ethereum network require
gas, a fraction of Ethereum token, which forms the justice and fairness
of the blockchain \cite{wood2014ethereum}.

\section{Notations}\label{notations}
In this session, we will introduce notations which were used in our
paper.
All participants involved in the vote can be divided into three types - 
\textit{voter}, \textit{organizer} and \textit{\queschain{} Contract}.

\begin{itemize}
    \item \textit{Voter}, one who has the permission to vote.
    \item \textit{Organizer}, one who initiated the vote, 
    each e-voting only has one organizer.
    \item \textit{\queschain{} Contract}, a Smart Contract on EVM, 
    which act as an inspector, contains
    \begin{itemize}
        \item Public key, 
        unchangeable key provided by the 
        organizer to check the signature.
        \item Judge function, a function to judge 
        if the ballot is valid or not.
        \item Ballot box a decentralized database to 
        store valid ballots.
    \end{itemize}
\end{itemize}
    
Let $\text{Voters}$ be the set of all the $\text{voter}$, 
\[
\forall \text{voter} \in \text{Voters}, \left| \text{Voters} \right|\ge 1.
\] 
Let $\text{Users}$ be the set of
$\text{organizer}$, $\text{\queschain{}\ Contract}$, and all
$\text{voter}$, 
\[
\text{Users}= \text{Voters} \cup \{\text{organizer}\} \cup \{\text{\queschain{}}\}.
\]
All the elements in $\text{Users}$ has its accounts in ETH.

Communications through ETH account during the e-voting, which will be
recorded by ETH, can be represented like, 
\[
a \xrightarrow[\text{memo}]{c}b\quad\text{where}\quad a,b \in \text{Users}
\]

and $c$ is the objects they what to send; $\text{memo}$ is the explanation authors
want to add.

Applying function $f$ to object $x$ can be represent like
$f(x)$,

e.g. 
\[
m'=\text{Hash}(m).
\]

Encrypt or decrypt message by function
$\text{Enc}(key,message)$ and $\text{Dec}(key,message)$.

e.g. 
\[
m'=\text{Enc}(pk,m)
\] 
where $m'$ is the encrypted message of
$m$ under key $pk$.

Data (i.e. Public key) owned by different types of participants by
Unified Modeling Language (UML) in Table \ref{table:participants}.

\begin{table}
\caption{Data Owned by Different Types of Participants}
\begin{center}
\begin{tabular}{|l|l|}
    \hline
    \textbf{organizer}     & \textbf{voter}           \\\hline
    $-$\verb|sk|                  & $-$\verb|r|                     \\\hline
    $+$\verb|pk|                  & $+$\verb|address|/$-$\verb|accounts|   \\\hline
    $+$\verb|address|/$-$\verb|accounts| & $-$\verb|address|/$-$\verb|accounts| \\\hline
                           & $-$\verb|uuid|                  \\\hline
\end{tabular}
\label{table:participants}
\end{center}
\end{table}

\begin{enumerate}
    \item \verb|sk|/\verb|pk|: to sign/check the signature.
    \item \verb|address|/\verb|accounts|: an ETH account, to prove who you are during communications.
    \item \verb|address|/\verb|accounts|`: an anonymous ETH account, which was randomly generated, destroyed immediately.
    \item \verb|r|: a key to blind the message, which was randomly generated, stored only in local.
    \item \verb|uuid|: to identify different ballots, which was randomly generated, stored only in local.
\end{enumerate}

\section{Details of Protocol}\label{details}

In this session, we will introduce each stage of our protocol during the
e-voting, which can be roughly divided into setup stage, sign stage, 
vote stage and count stage.

\subsection{Setup Stage}

The organizer should initialize the e-voting by following steps,

\begin{enumerate}
\item
  construct a \queschain{} like Smart Contract 
  with its $pk$, vote-start time $st$, 
  vote-check time $ct$ and vote-end time $et$.
  \[
  \text{organizer} \xrightarrow[\text{unchangable}]{pk,st,ct,et}\text{\queschain{}}
  \]
\item
  construct and publish the voting page.
\item
  make a permission list of all voters' address like, \[
  \text{Voters}=\{\text{voterA}, \text{voterB},\text{voterC},\ldots\}
  \] Applying function $\text{Chance}()$ on voter to find out number of
  submissions the voter allowed.
\end{enumerate}

Then the voters can construct their ballots, represented by
$\text{ballot}$.

\begin{figure}
\centerline{\includegraphics[width=0.5\textwidth]{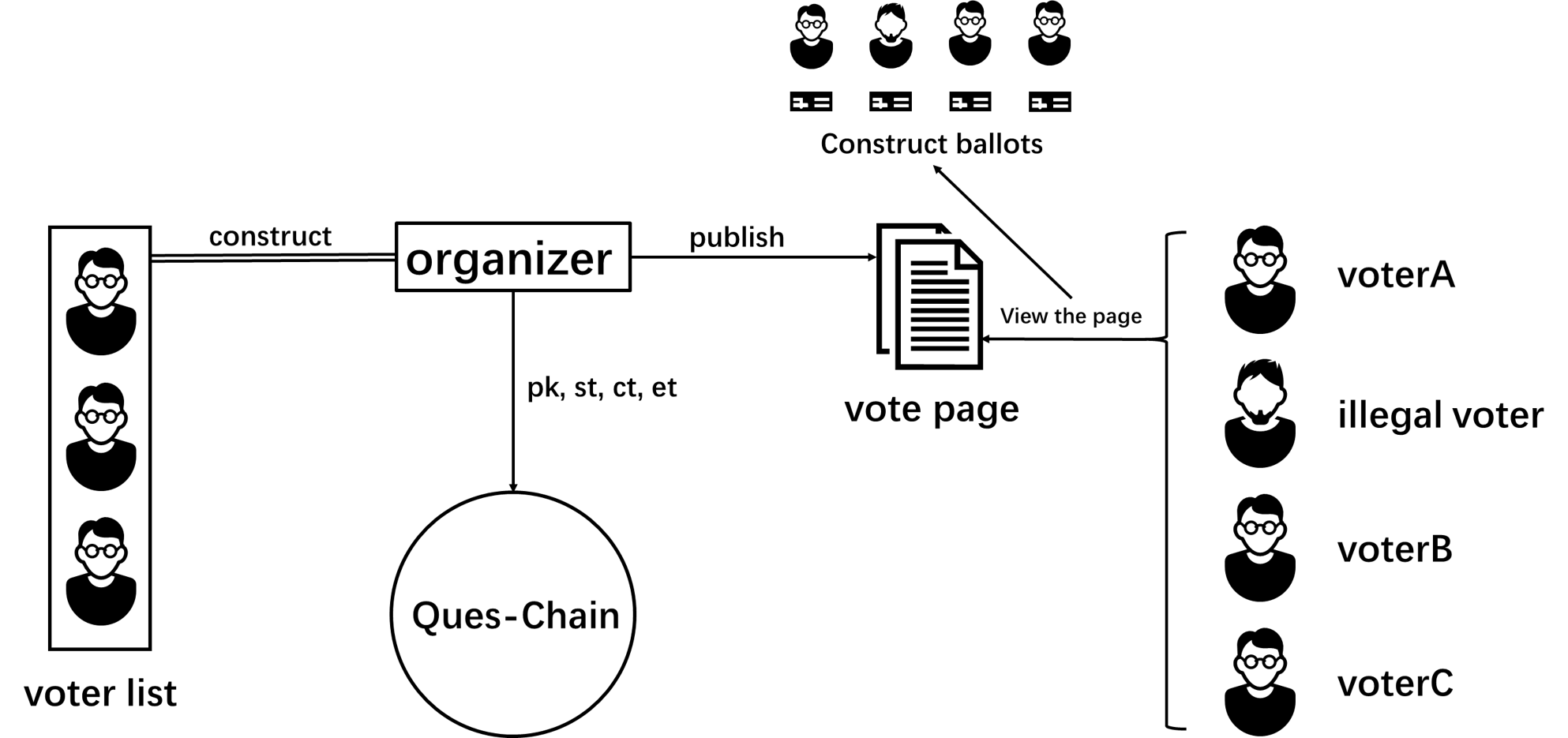}}
\caption{Setup stage in \queschain{} protocol.}
\label{fig:protocol}
\end{figure}

\subsection{Sign Stage}

In this stage, the voter will get a signed-blinded-ballot from the
organizer. Before $ct$ and after $st$, the voter should do
the following steps.

Firstly, the voter need to get the hash of $\text{ballot}$.

Then, the voter randomly generated a key $r$ and an $\text{uuid}$, 
stored only locally. Append the $\text{uuid}$ into the end of $m$ and
encrypt them by $r$, get the $\text{BlindedBallot}$. \[
\text{BlindedBallot}=\text{Enc}(r,\text{Hash}(\text{ballot})+\text{uuid})
\] where the operator $+$ means append into the end.

After that, the voter send $\text{BlindedBallot}$ to the organizer through 
its ETH accounts which had the permission to vote. \[
\text{voter} \xrightarrow{\text{BlindedBallot}}\text{organizer}
\]
The organizer could decide whether to sign or not by Algorithm \ref{algo:1}
and then send the $\text{SignedBlindedBallot}$ to the voter. 
\[
\text{organizer}\xrightarrow{\text{SignedBlindedBallot}}\text{voter}
\] It should be noted that the organizer can't get any information of
$r$ and $\text{uuid}$ for them only being stored locally.

\begin{figure}
  \removelatexerror
  \begin{algorithm}[H]
    \caption{Decide whether to sign or not.}\label{algo:1}
    \begin{algorithmic}[1]
      \REQUIRE address, BlindedBallot
      \IF{address in Voters \AND Chance(address)>0}
      \STATE SignedBlindedBallot = Enc(sk, BlindedBallot)
      \STATE Chance(address) = Chance(address) - 1
      \ELSE
      \STATE SignedBlindedBallot = 0
      \ENDIF
      \RETURN SignedBlindedBallot
    \end{algorithmic}
  \end{algorithm}
\end{figure}

And for the voter, to ensure the organizer give all the voters the same
signature, he should send $\text{SignedBlindedBallot}$ and
$\text{BlindedBallot}$ to \queschain{} Contract to check the signature.

In the end, the voter send $\text{SignedBlindedBallot}$ and $\text{BlindedBallot}$ 
to \queschain{} Contract for checking.
\[
\text{voter}\xrightarrow[\text{SignedBlindedBallot}]{\text{BlindedBallot}}\text{\queschain{}}
\]

\queschain{} Contract will check the signature and return $\text{result}$
by Algorithm \ref{algo:2}.

\begin{figure}
  \removelatexerror
  \begin{algorithm}[H]
    \caption{Signature check in \queschain{} contract.}\label{algo:2}
    \begin{algorithmic}[1]
      \REQUIRE SignedBlindedBallot, BlindedBallot
      \IF{Dec(pk, SignedBlindedBallot) = BlindedBallot}
      \STATE result = True
      \ELSE
      \STATE result = False
      \ENDIF
      \RETURN result
    \end{algorithmic}
  \end{algorithm}
\end{figure}

If $\text{result}$ is True, the voter can step into next stage,
else he should communicate with the organizer to find out what's wrong
with the $\text{SignedBlindedBallot}$ or $\text{BlindedBallot}$.

\begin{figure}
  \centerline{\includegraphics[width=0.5\textwidth]{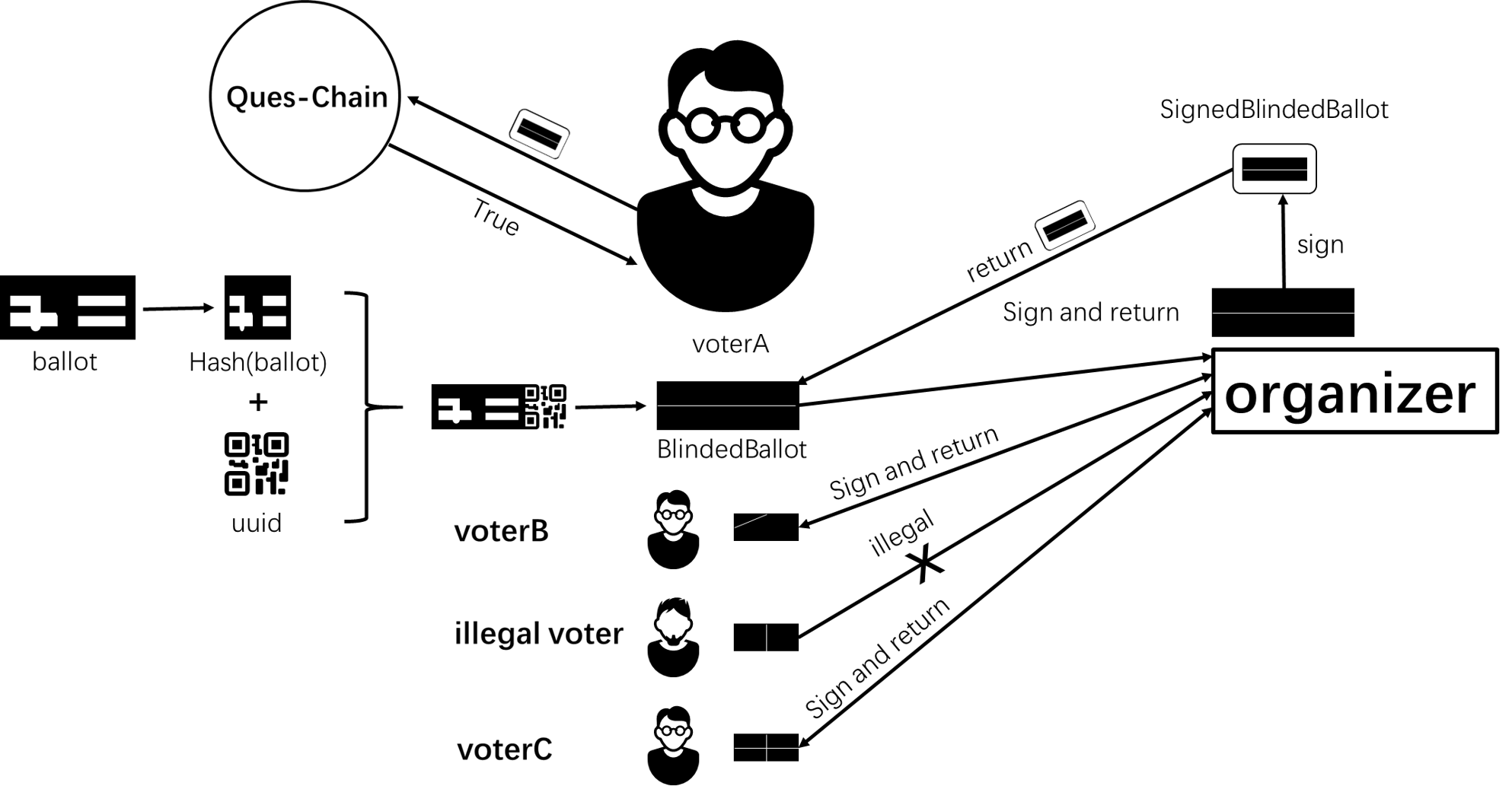}}
  \caption{Sign stage in \queschain{} protocol}
  \label{fig:contract}
\end{figure}

\subsection{Vote Stage}
In this stage, the voter will vote. Before $et$ and after
$ct$, the voter should do the following steps.

\begin{enumerate}
\item
  decrypt the $\text{SignedBlindedBallot}$ to $\text{SignedBallot}$. 
  \[
  \text{SignedBallot}=\text{Dec}(r,\text{SignedBlindedBallot})
  \]
\item
  randomly generated a new ETH account
  $\text{account}'$ with
  $\text{address}'$, stored only locally.
\item
  anonymously send the $\text{SignedBallot}$, $\text{Ballot}$,
  $\text{uuid}$ to \queschain{} Contract through
  $\text{account}'$. \[
  \text{voter}\xrightarrow[\text{through account}']{\text{SignedBallot},\text{Ballot},\text{uuid}}\text{\queschain{}}
  \]
\end{enumerate}

It should be noted that $\text{account}'$ is generated
randomly and only stored locally, so no one (except the voter) will know
whom $\text{account}'$ belongs to, which made it
untraceable.

\subsection{Count Stage}
In this stage, the voter will count legal ballots and publish the
result.

Here we use a map $\text{BallotBox}$ to store legal ballots and
its uuid. Each time \queschain{} Contract receive a ballot, \queschain{}
Contract will check the ballot by Algorithm \ref{algo:3}.

\begin{figure}
  \removelatexerror
  \begin{algorithm}[H]
    \caption{Judge function.}\label{algo:3}
    \begin{algorithmic}[1]
      \REQUIRE SignedBallot,Ballot,uuid
      \IF{Dec(pk,SignedBallot) = Hash(Ballot) + uuid \\
      \AND \NOT (uuid in BallotBox.values())}
      \STATE Add map (Ballot, uuid) into BallotBox
      \STATE result = True
      \ELSE
      \STATE result = False
      \ENDIF
      \RETURN result
    \end{algorithmic}
  \end{algorithm}
\end{figure}

Every legal ballot will be stored in $\text{BallotBox}$ while illegal
ballots will be ignored.

After $et$, all users of the ETH can get the result of the vote
by counting all ballots in $\text{BallotBox}$.

\subsection{Publish Option}
Due to the feature of ETH, the result of the vote will be published on
ETH and everyone which everyone is accessible. However, in some scenery,
the organizer may want to keep secret of the result or want to control
whether push it. Here we give an option for the organizer to set whether
publish or not by generating an extra
$pk''-sk''$
and use the $\text{Enc}(pk,\text{Ballot})$ replaces the $\text{Ballot}$ on sign
stage and vote stage. The organizer can publish the result by publishing
$sk''$.

\section{Security Analysis}\label{security}

According to the standard mentioned on the reference, our protocol
equipped following security properties:

\textit{Correctness} If all the election's participants, such as voters,
authorities and so on are honest and behave as it is scheduled, then the
final results are effectively the tally of casted votes.

\textit{Privacy} No participant other than a voter should be able to
determine the value of the vote cast by that voter.

\textit{Robustness} Faulty behavior of any coalition of authorities can
be tolerated. No coalition of voters can disrupt the election, and any
cheating voter will be detected.

\textit{Verifiability} Correct voting processes must be verifiable to
prevent incorrect voting results.

\textit{Democracy} There are two requirements to satisfy in democracy,
\begin{itemize}
  \item \textit{Eligibility}: only authorized voters are allowed to vote.
  \item \textit{Prevention of multiple voting}: all eligible voters are allowed to cast
    the scheduled vote's number (function of the election system and his 
    part in it) and not more, such that each voter has his intended power
    in deciding the outcome of the voting.
\end{itemize}

\textit{Fairness} No participant can gain any knowledge, except his
vote, about the (partial) tally before the counting stage (The knowledge
of the partial tally could affect the intentions of the voters who has
not yet voted.)

However, some attacks remain as follow.

\textit{Receipt-Freeness} Voters must neither be able to obtain nor
construct a receipt which can prove the content of their vote.

For the voters have key $r$, which was used to blind the ballots,
to prove the content of their vote, our protocol doesn't equip
Receipt-Freeness.

In general, the security properties of the protocol as shown in Table \ref{table:security}.

\begin{table}
  \caption{Security Properties of the Protocol}
  \begin{center}
  \begin{tabular}{|l|l|c|}
      \hline
      \multicolumn{2}{|c|}{\textbf{Requirement}} & \textbf{Property} \\ \hline
      \multicolumn{2}{|c|}{Privacy} & Correct \\ \hline
      \multicolumn{2}{|c|}{Receipt-Freeness} & Attacks Found \\ \hline
      \multicolumn{2}{|c|}{Robustness} & Correct \\ \hline
      Verifiability & \multicolumn{1}{|@{}c@{}|}
      {\begin{tabular*}{4em}{l}
        U \\ \hline
        I
      \end{tabular*}} & Correct \\ \hline
      Democracy & \multicolumn{1}{|@{}c@{}|}
      {\begin{tabular*}{4em}{l}
        E\quad \\ \hline
        PMV
      \end{tabular*}} & Correct \\ \hline
      \multicolumn{2}{|c|}{Fairness} & Correct \\  \hline
      \multicolumn{2}{|c|}{Correctness} & Correct \\ \hline
  \end{tabular}
  \label{table:security}
  \end{center}
  \end{table}

\section{Usages}\label{usages}

\queschain{} can be applied to applications varying from the national
referendum to internal evaluations conducted by companies.

In the case of election or referendum, \queschain{} guarantees the
consistency of the rule for all voters, making sure that every voter has
chances exactly the organizer given to vote. Computing tasks can be
distributed either throughout the Ethereum main network or to everyone
in the country of election who volunteers to verify the computation with
his or her computer.

The estimated cost of conducting the Brexit referendum using \queschain{}
on the Ethereum main network is 3.9 million pounds, 97\% lower than the traditional
way which cost 129.1 million pounds \cite{electoral2018costs}. 
The latter way of
using volunteer computation power may save even more. Online retailers
may utilize \queschain{} to collect customers' reviews. With full control
of submission permission, shopping sites save money and time countering
spam reviews, being able to show real ones to interested customers.
Corporations can conduct employee evaluations using \queschain{}.
Anonymity is protected to encourage real feedback and prevent gossips or
bullying. Non-governmental organizations (NGOs) may also evaluate projects 
they carried out or sponsored with the help of \queschain{}.

\queschain{} can be used in a situation which has a high
information-security requirements. For the organizations want to hear
real thoughts from their employees and clients. However, the latter may have
concerns about whether their expression will bring them bad
consequences. Such concerns may lead to distorted feedback results. Even
if organizations don't care about the exact identity of feedback
sources, with current technology, they are not equipped to prove or ensure
that they won't do so.

Votes, questionnaires, outcome assessment, opinion collection, complaint
reporting, etc., all applications which involve anonymous feedback can
take advantage of \queschain{}. While the permission right is held by
organizers, voters' anonymity is still protected.

\begin{figure}
    \centerline{\includegraphics[width=0.5\textwidth]{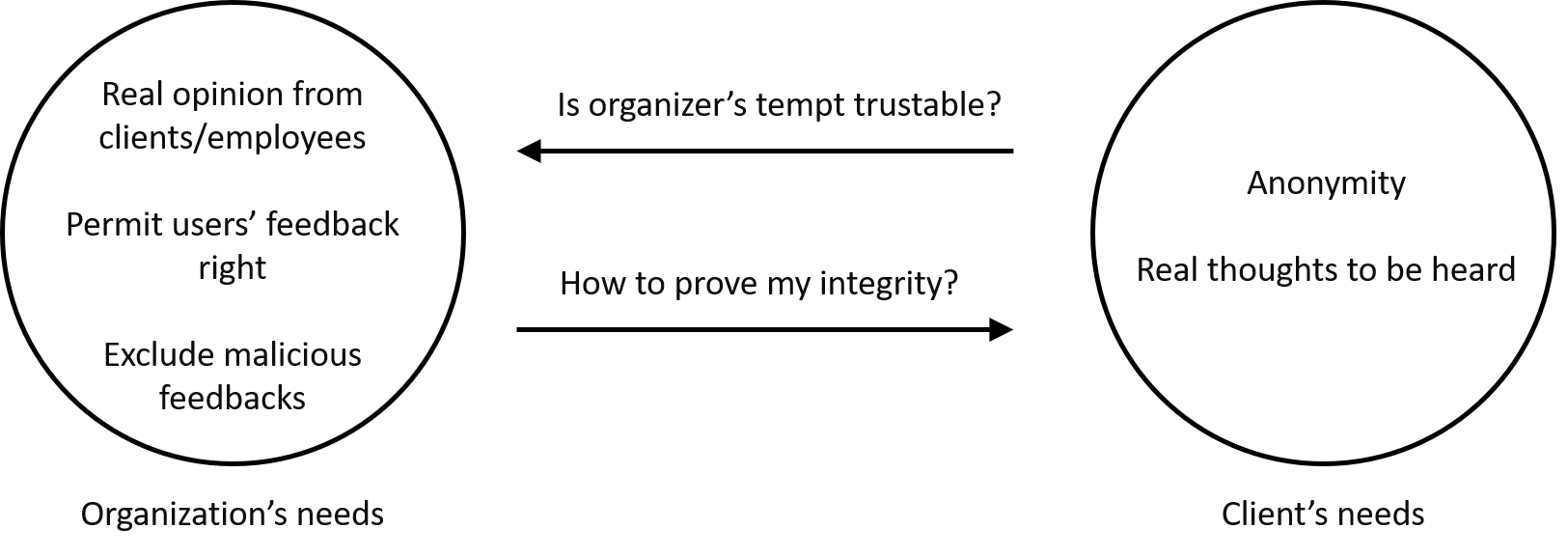}}
    \caption{Applications of \queschain{} protocol.}
    \label{fig:usage}
  \end{figure}


\end{document}